\documentclass[useAMS,usenatbib,usegraphicx]{mn2e}

\usepackage{natbib}
\usepackage{graphicx}
\usepackage{amsmath}
\usepackage{amssymb}
\usepackage{color}

\newcommand {\ie} {{i.e.}}

\newcommand {\eg} {{e.g.}}
\newcommand {\ea} {{et~al.}}
\newcommand {\be} {\begin{equation}}
\newcommand {\ee} {\end{equation}}

\newcommand {\p} {\partial}
\newcommand {\beq} {\begin{equation}}
\newcommand {\eeq} {\end{equation}}

\newcommand {\ds} {\displaystyle}

\title{Application of the spine-layer jet radiation model to outbursts in the broad-line radio galaxy 3C~120}

\author[Janiak \ea]{M.~Janiak$^{1}$\thanks{E-mail: mjaniak@camk.edu.pl}, M.~Sikora$^{1}$, and R.~Moderski$^{1}$ \\ $^{1}$Nicolaus Copernicus Astronomical Center, Bartycka 18, 00-716 Warsaw, Poland }

\begin{document} 

\maketitle

\begin{abstract}
We present a detailed {\it Fermi}/LAT data analysis for the broad-line radio galaxy 3C~120. This source has recently entered into a state of increased $\gamma$-ray activity which manifested itself in two major flares detected by {\it Fermi}/LAT in September 2014 and April 2015 with no significant flux changes reported in other wavelengths. We analyse available data focusing our attention on aforementioned outbursts. We find very fast variability timescale during flares (of the order of hours) together with a significant $\gamma$-ray flux increase. We show that the $\sim6.8$~years averaged $\gamma$-ray emission of 3C~120 is likely a sum of the external radiation Compton and the synchrotron self-Compton radiative components. To address the problem of ``orphan'' $\gamma$-ray flares and fast variability we model the jet radiation dividing the jet structure into two components: the wide and relatively slow outer layer and the fast, narrow spine. We show that with the addition of the fast spine occasionally bent towards the observer we are able to explain observed spectral energy distribution of 3C~120 during flares with the Compton up-scattered broad-line region and dusty torus photons as main $\gamma$-rays emission mechanism.
\end{abstract}


\section{Introduction}
The recent detection of $\gamma$-rays from several broad-line radio galaxies (BLRGs: 3C~120, 3C~111, and Pictor~A) by {\it Fermi}/LAT telescope (\citealp{Abdo10b}; \citealp{Kat11}; \citealp{Bro12}) constituted this sub-class of active galactic nuclei (AGN) as high energy emitters. Together with radio galaxies such as  M87 \citep{Abdo09} or Cen~A \citep{Abdo10} seen by {\it Fermi}/LAT earlier they form an interesting group of AGN where high energy radiation is observed despite the absence of strong Doppler boosting as we observe in typical blazars. These non-blazar AGN (dubbed ``misaligned blazars''), although much fainter due to geometrical effects, are very interesting objects as they reveal both accretion disk radiation (at least in case of BLRGs) and jet emission in their broadband spectra. 

From infrared to hard X-rays BLRGs show typical thermal emission related to accretion disk around super-massive black hole, which can be further divided into direct accretion disk radiation peaking in optical wavelengths, dusty torus infrared emission and power law non-thermal X-ray component most probably originating from disk corona (\citealp{Zdzia01}, \citealp{Gra07}). Non-thermal radio and $\gamma$-ray emission is thought to originate from relativistic jet and is being only weakly Doppler boosted due to large viewing angles $\theta_{\mathrm{obs}} \gtrsim 10\degr$.

3C~120 is a nearby ($z = 0.033$) BLRG with Fanaroff-Riley type I radio morphology \citep{Wal87}. The source has been actively monitored in all wavebands from radio up to X-rays and recent detection by {\it Fermi}/LAT \citep{Abdo10} made it possible to study accretion disk and jet interaction by modelling its broadband spectra. 

In this article we present detailed data analysis of high energy data for the whole {\it Fermi}/LAT dataset ($\sim6.8$~years) and we especially focus our attention on two major outbursts that happened in September 2014 and April 2015. We study both temporal and spectral properties of these flares.

This paper is organised as follows. In section~\ref{sec:data} we present the details and results of {\it Fermi}/LAT data analysis. In section~\ref{sec:dis} we present discussion concerning the location of the $\gamma$-ray emission region (the ``blazar zone''), the adopted ``spine-layer'' jet model as well as details on spectral modelling. Summary and conclusions are presented in section~\ref{sec:con}.

\section{{\it Fermi}/LAT data analysis}
\label{sec:data}
The 3C~120 $\gamma$-ray spectra and light curves within~100~MeV to 100~GeV energy range were obtained by analysing about~$6.8$~years of {\it Fermi}/LAT data from August~4th, 2008 to May~26th, 2015 (hereafter: all {\it Fermi}/LAT data). We reduced the data using {\it Fermi} Science Tools package version \verb|v9r33p0| selecting only \verb|source| class 2 events  and using \verb|P7REP_SOURCE_V15| instrument response function. Given a rather soft spectrum of 3C~120 and a large instrument point spread function ($3.5\degr$ at $100$~MeV) we selected events within $20\degr$ (ROI) from nominal source radio position (R.A.=$68.296\degr$, Dec.=$5.354\degr$) to properly model the contribution of other sources. To avoid contamination form Earth's limb photons we also applied a zenith angle~$>100\degr$ cut.

We used the standard \verb|gtlike| tool to model the ROI using maximum likelihood method \citep{Cash79}. Model includes 69 point sources from LAT 4-year Point Source Catalog (3FGL, \citealp{Ack15}) as well as Galactic and isotropic diffusion emission templates \footnote{\texttt{gll\_iem\_v05\_rev1.fit} and \texttt{iso\_source\_v05.txt}, respectively.}. We model both flux normalisation and power law spectral index for sources within a radius of $10\degr$ from the ROI centre. Other sources within the ROI are fixed to their 3FGL values. After initial fit, all sources with test statistics (TS) lower than~$1.0$ were excluded from the model and the procedure was repeated until convergence.

Two sources in the 3FGL are located fairly close to 3C~120~\ie~3FGL J0432.5+0539 and 3FGL J0426.6+0459. The latter is $1.66\degr$ away yet the former is only $0.35\degr$ away from the source of interest having photon index of $\alpha=2.7$ which makes distinction between the two sources unclear. However, by fitting $\gamma$-ray 3C~120 position \citet{Tan15} found that it lies $0.028\degr \pm 0.088\degr$ from its radio position. Also, 3FGL J0432.5+0539 LAT position was determined with a $95\%$ error of $0.15\degr$. Therefore, we include all three sources in the model treating them as separate.

For the whole dataset (all data) we used binned likelihood method. 3C~120 was detected with $\mathrm{TS}=156$, a $100\,\textrm{MeV}-100\,\textrm{GeV}$ flux of $F=(3.0 \pm 0.6) \times 10^{-8}~\textrm{ph}~\textrm{cm}^{-2}~\textrm{s}^{-1}$ and a power law photon index $\alpha=2.71 \pm 0.09$. We calculated light curve and spectra by dividing the data into time bins (30-days, 83 bins) or energy bins (6 equal bins in logarithmic scale covering the chosen energy range), and by applying previous procedure to model the ROI with both flux normalisation and power law index as free parameters. In case the TS value was lower than 9 (corresponding to significance $\sigma \sim 3$) $95\%$ upper limits were calculated assuming power law index of $\alpha=2.7$. Figure~\ref{fig:base_spectra} presents source $\gamma$-ray spectra.

\setcounter{figure}{0}
\begin{figure*}
  \includegraphics[width=.6\textwidth]{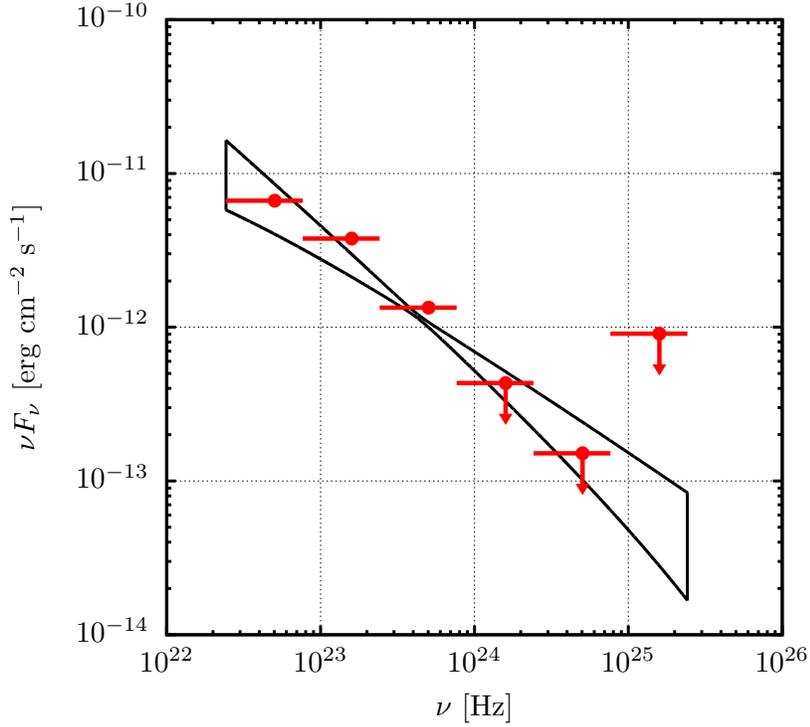}
 \caption{3C~120 spectra in $100\,\textrm{MeV}-100\,\textrm{GeV}$ energy range for the whole dataset from 4th August 2008 to 26th May 2015. Arrows indicate $95\%$ upper limits for detection with significance lower than $3\sigma$.}
  \label{fig:base_spectra} 
\end{figure*}

\setcounter{figure}{1}
\begin{figure*}
  \includegraphics[width=.6\textwidth]{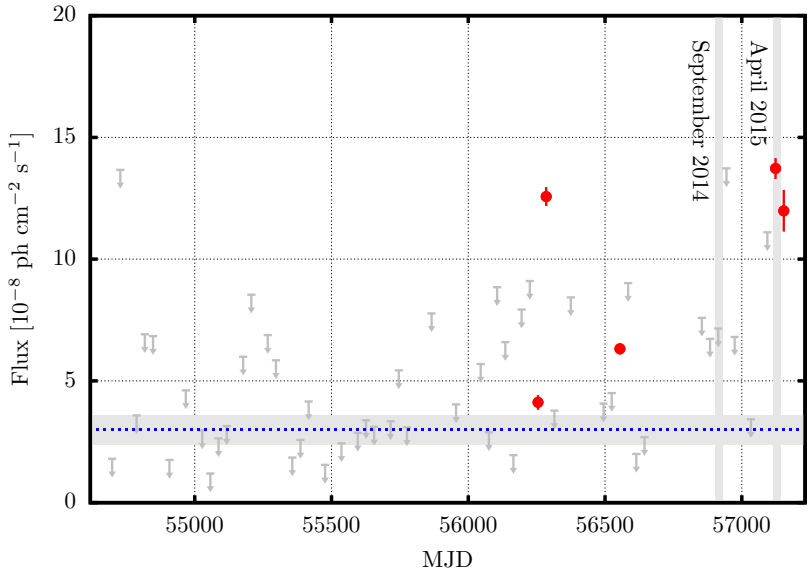}
 \caption{3C~120 light curve ($100\,\textrm{MeV}-100\,\textrm{GeV}$) from 4th August 2008 to 26th May 2015 with 30-days time bins. Red points correspond to detection above $3\sigma$ significance while grey points indicate $95\%$ upper limits for detection with significance lower than $3\sigma$. Blue, dotted line and light-blue shaded area correspond to $6.8$~years average {\it Fermi}/LAT flux and flux error, respectively.}
  \label{fig:base_lc} 
\end{figure*}

\setcounter{figure}{2}
\begin{figure*}
  \includegraphics[width=.45\textwidth]{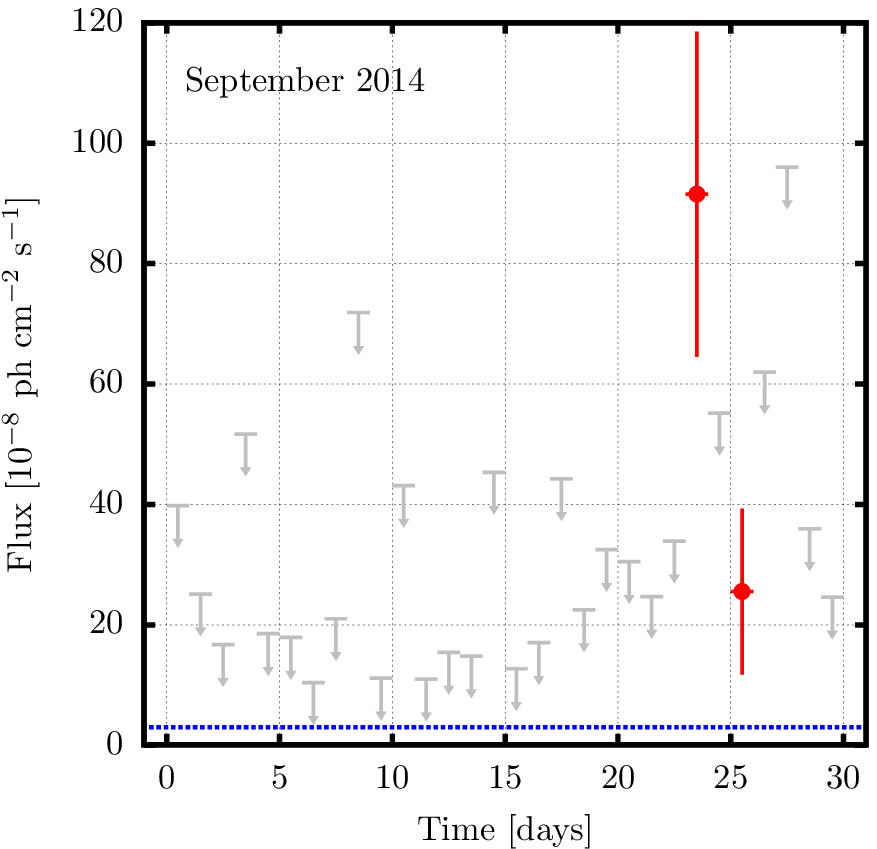}
  \includegraphics[width=.45\textwidth]{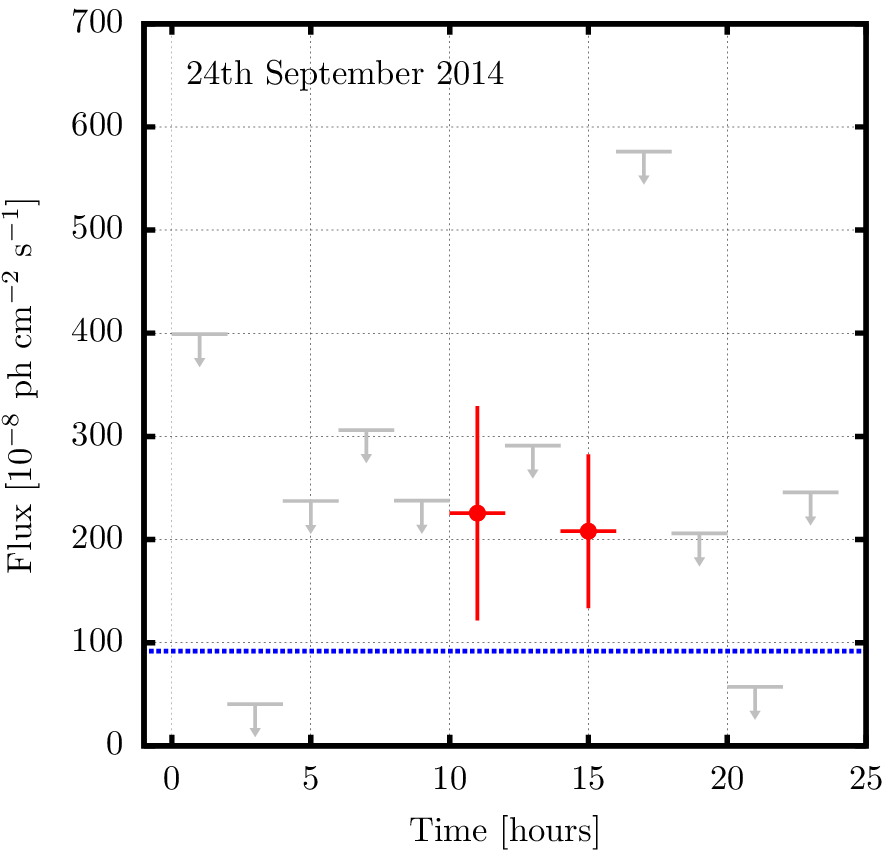}
 \caption{3C~120 light curve ($100\,\textrm{MeV}-100\,\textrm{GeV}$). \emph{Left panel:} data for September 2014 with 1-day time bins. \emph{Right panel:} data for 24th September 2014 with 2-hour time bins. Red points correspond to detection above $3\sigma$ significance while grey points indicate $95\%$ upper limits for detection with significance lower than $3\sigma$. Blue, dotted line correspond to monthly (\emph{left panel}) and daily (\emph{right panel}) average {\it Fermi}/LAT flux.}
  \label{fig:flare14_lc} 
\end{figure*}

\setcounter{figure}{3}
\begin{figure*}
  \includegraphics[width=.45\textwidth]{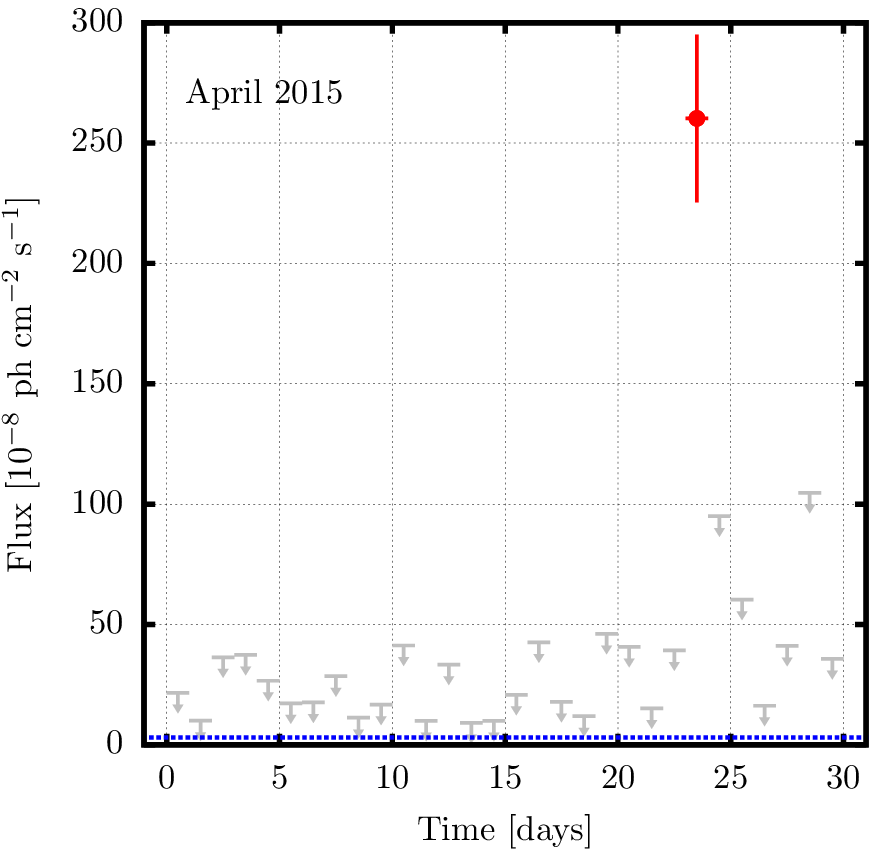}
  \includegraphics[width=.45\textwidth]{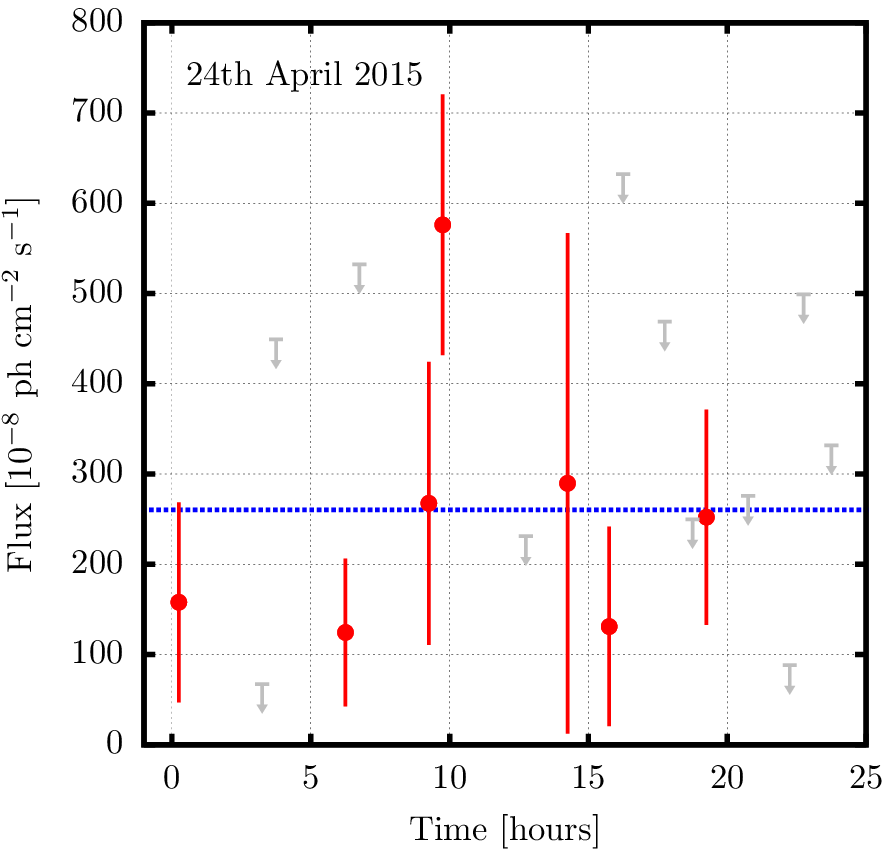}
 \caption{3C~120 light curve ($100\,\textrm{MeV}-100\,\textrm{GeV}$). \emph{Left panel:} data for April, 2015 with 1-day time bins. \emph{Right panel:} data for 24th April 2015 with 30-minutes time bins. Red points correspond to detection above $3\sigma$ significance while grey points indicate $95\%$ upper limits for detection with significance lower than $3\sigma$. Blue, dotted line correspond to monthly (\emph{left panel}) and daily (\emph{right panel}) average {\it Fermi}/LAT flux.}
  \label{fig:flare15_lc} 
\end{figure*}

Figure~\ref{fig:base_lc} presents 3C~120 light curve. Source was observed with significance larger that $3\sigma$ several times in recent years. We focus our attention on two recent outbursts: in September 2014 (hereafter: flare 1; reported by \citealp{Tan14atel}) and in April 2015 (hereafter: flare 2; reported by \citealp{FLatTeam}). To analyse {\it Fermi}/LAT data during flare 1 and 2 periods we used unbinned likelihood analysis method to properly account for low photon count numbers. During flare 1 in September 2014 (30-days of data) 3C~120 was observed with $\mathrm{TS}=11$, power law photon index $\alpha=2.50 \pm 0.32$ and its $95\%$ flux upper limit was estimated as $F_\mathrm{UL} = 1.2 \times 10^{-7}~\textrm{ph}~\textrm{cm}^{-2}~\textrm{s}^{-1}$. Figure~\ref{fig:flare14_lc} (left panel) presents light curve for September 2014 with clear flux increase on 24th September. Right panel of the same figure presents light curve for that day with 2-hour time bins. On 24th September 3C~120 was detected with $\mathrm{TS}=54$ with photon flux $F = (9.2 \pm 2.7 ) \times  10^{-7}~\textrm{ph}~\textrm{cm}^{-2}~\textrm{s}^{-1}$ and power law index of $\alpha = 2.34 \pm 0.25$. Even for the 2-hour highest flux time bin $\mathrm{TS}=28$ indicating detection above $5\sigma$ significance level.

We applied the same procedure to investigate flare 2. Figure~\ref{fig:flare15_lc} presents monthly (1-day time bin) and daily (30-minutes time bin) light curves for April 2015 and 24th April 2015, respectively. Monthly averaged flux is $F = (1.45 \pm 0.09 ) \times  10^{-7}~\textrm{ph}~\textrm{cm}^{-2}~\textrm{s}^{-1}$ with power law index $\alpha = 2.47 \pm 0.03$ ($\mathrm{TS} = 57$) and daily averaged flux for 24th April is $F = (2.6 \pm 0.35 ) \times  10^{-6}~\textrm{ph}~\textrm{cm}^{-2}~\textrm{s}^{-1}$ with power law index $\alpha = 2.22 \pm 0.11$ ($\mathrm{TS} = 281$). Source was also detected within 30-minutes time bin with large $\mathrm{TS}=89$.

Figure~\ref{fig:flare_spectra} presents flares 1~and~2 $\gamma$-ray spectra averaged over outburst day and shortest time bin with detection above $~3\sigma$ i.e. 2 hours for flare 1 and 30 minutes for flare 2. 

\setcounter{figure}{4}
\begin{figure*}
  \includegraphics[width=.45\textwidth]{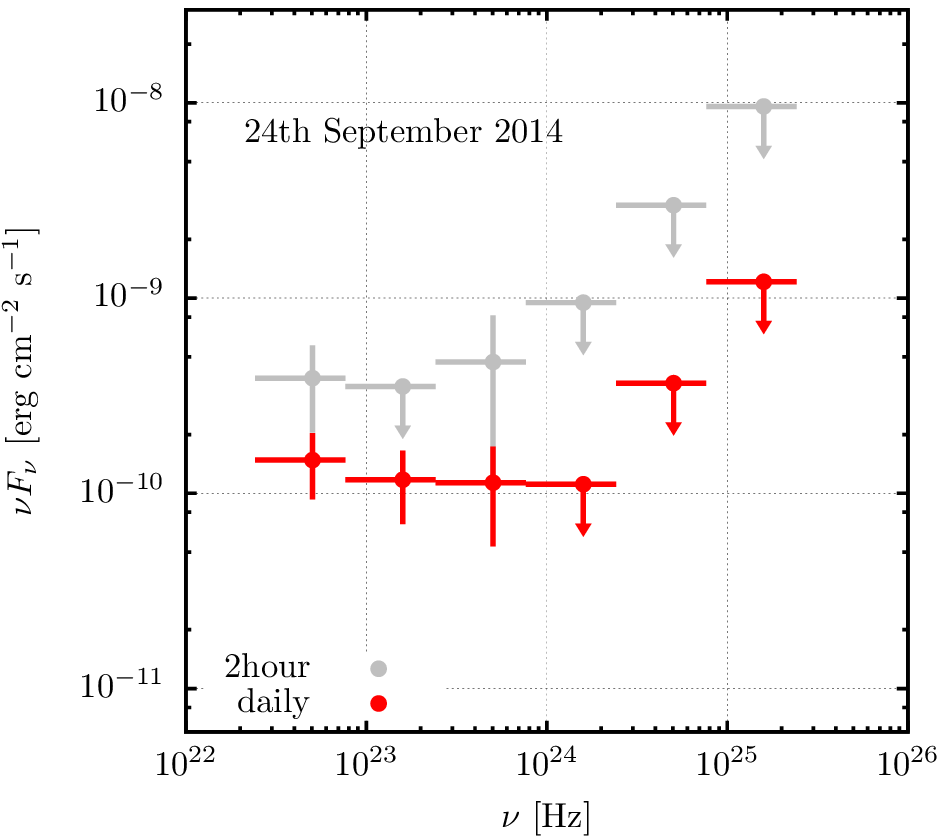}
  \includegraphics[width=.45\textwidth]{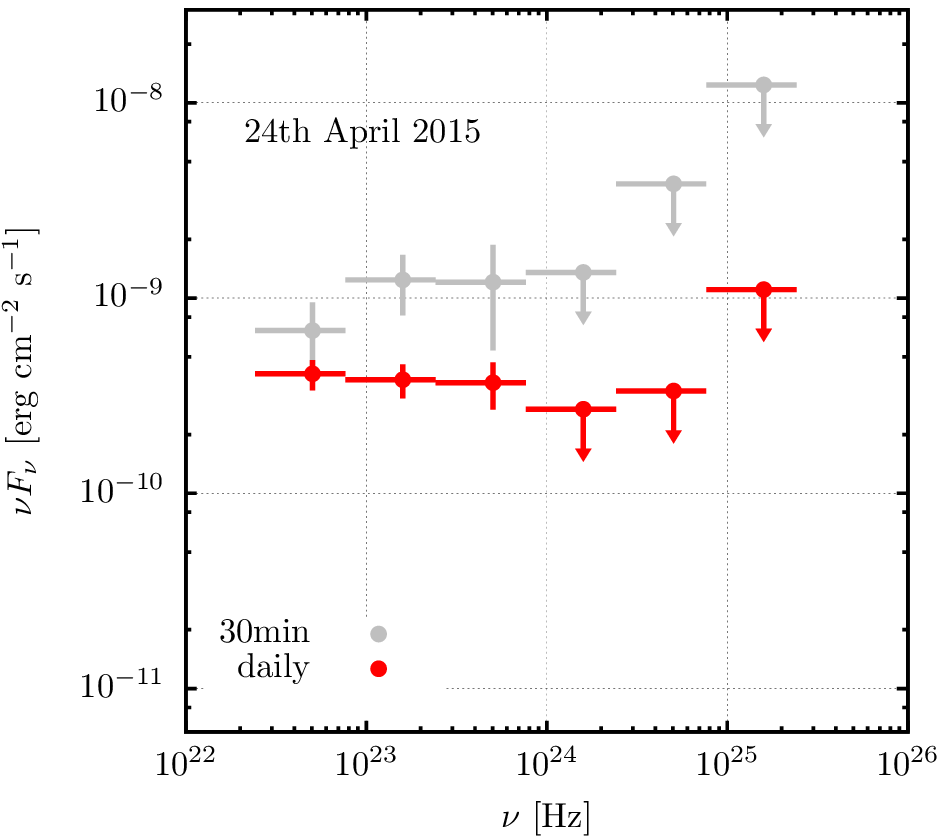}
 \caption{3C~120 spectra in $100\,\textrm{MeV}-10\,\textrm{GeV}$ energy range for flare 1 (\emph{left panel}) and flare 2 (\emph{right panel}). Arrows indicate $95\%$ upper limits for detection with significance lower than $3\sigma$. Red points present daily averaged spectra while grey points show spectra for highest flux 2-hours time bin (flare 1) and 30-minutes time bin (flare 2).}
  \label{fig:flare_spectra} 
\end{figure*}

For flaring periods we additionally checked whether they could be associated with nearby sources 3FGL J0432.5+0539 and 3FGL J0426.6+0459. Light curve analysis for these sources indicated no correlation between their flux increase and presented 3C~120 outbursts. 

Table~\ref{tab:data} summarises {\it Fermi}/LAT data analysis results.

\setcounter{table}{0}
\begin{table*}
\caption{Summary of {\it Fermi}/LAT data analysis results for 3C~120. $\alpha$ and $\mathrm{TS}$ denote power law photon index and test significance, respectively.}
\begin{center}
\begin{tabular}{p{0.25\textwidth}@{}r p{0.1\textwidth}@{}r p{0.1\textwidth}@{}r p{0.1\textwidth}r}
\hline
\hline
Time period & Flux [$10^{-7} \textrm{ph}~\textrm{cm}^{-2}~\textrm{s}^{-1}$] & $\,\,\,\,\,\,\,\,\,\,\,\alpha$ & $\mathrm{TS}$ \\
\hline
$6.8$ years & $0.3 \pm 0.06$ & $2.71 \pm 0.09$ & $156$ \\
\hline
Flare 1 & & & \\
September 2014 & $1.2$ ($95\%$ upper limit) & $2.50 \pm 0.32$ & $11$ \\
24th September 2014 & $9.2 \pm 2.7$ & $2.34 \pm 0.25$ & $54$ \\
2-hour highest flux time bin & $21 \pm 7.5$ & $2.40 \pm 0.35$ & $28$ \\
\hline
Flare 2 & & & \\
April 2015 & $1.45 \pm 0.09 $ & $2.47 \pm 0.03$ & $57$ \\
24th April 2015 & $26 \pm 3.5$ & $2.22 \pm 0.11$ & $281$ \\
30-minutes highest flux time bin & $58 \pm 14$ & $2.07 \pm 0.21$ & $89$ \\
\hline
\hline
\end{tabular}
\end{center}
\label{tab:data}
\end{table*}

\section{Discussion}
\label{sec:dis}
\subsection{Location of the blazar zone}
\label{sec:location}
{\it Fermi}/LAT data suggest that 3C~120 underwent very rapid changes in $\gamma$-ray flux. Both aforementioned flares happened on very short timescales. Assuming that characteristic variability timescale $t_{\mathrm{var}}$ equals the shortest time bin for which the source has been detected with significance larger than $3\sigma$ we estimate $t_{\mathrm{var}}=2\,\mathrm{h}$ for flare~1 and $t_{\mathrm{var}}=30\,\mathrm{min}$ for flare~2. In a similar way 3C~120 $\gamma$-ray variability timescale for six years of {\it Fermi}/LAT data was estimated to be $5-10$~days by~\cite{Tan15}.

Variability timescale and causality arguments allow us to estimate radius of the emission region $R \leq c \mathcal{D} t_{\mathrm{var}}$ where $\mathcal{D} = [\Gamma (1-\beta\cos{\theta_{\mathrm{obs}}})]^{-1}$ is the Doppler factor, $\Gamma$ is the jet Lorentz factor, $\beta$ is the jet velocity in the speed of light units and $\theta_{\mathrm{obs}}$ is the angle towards the observer. For 3C~120 the average value of $\Gamma = 5.3 \pm 1.2$ \citep{Jor05a} which is in agreement with further work by \citet{Cas15}. The angle towards the observer $\theta_{\mathrm{obs}}$ was estimated in several papers using VLBI observations to have different mean values: from $9.7\degr$ \citep{Hov09} to $20.5\degr$ \citep{Jor05a} indicating that the jet might be changing its directions or that the radio knots might be moving not along the jet axis. For further considerations we adopt a mean value of $\theta_{\mathrm{obs}} = 15\degr$. Assuming aforementioned variability timescales and the conical jet opening angle $\theta_{\mathrm{jet}} = 1/\Gamma$ we estimate the radius of the emitting region to be $R \lesssim 1.0 \times 10^{17} (\mathcal{D}/5.3)$~cm for all data, $R \lesssim 1.1 \times 10^{15}(\mathcal{D}/5.3)$~cm for flare 1 and $R \lesssim 2.9 \times 10^{14}(\mathcal{D}/5.3)$~cm for flare 2. 

Using $\Gamma=5.3$ and $\theta_{\mathrm{obs}}=15\degr$ which results in $\mathcal{D}=3.7$, in agreement with values reported in the literature (\citealp{Hov09}; \citealp{Jor05a}; \citealp{Cas15}), we estimate the location of $\gamma$-ray emission region (the blazar zone) from the central black hole (BH) at $r = \Gamma R \lesssim 5.3 \times 10^{17}$~cm~$\approx 0.17$~pc for all {\it Fermi}/LAT data. Similar estimates were calculated by other works. \citet{Chatt09} calculated a de-projected distance of $\sim0.5$~pc from the central BH to the $43$ GHz VLBA radio core by using the time lag between the dip in the X-ray flux (assumed to originate in accretion disc corona) and the ejection of superluminal component from the radio core. By analysing {\it Fermi}/LAT lighcurves and radio data \citet{Cas15} found that $\gamma$-ray emission region lies about $\sim 0.13$~pc upstream from the radio core. Similar conclusions were drawn by \citet{Tan15} locating the emission region at $0.1-0.3$~pc from the central BH.

By assuming that the $\gamma$-ray emission region is transversly uniform relation $r = \Gamma R$ gives the location of the blazar zone at $r \lesssim 5.8 \times 10^{15}(\mathcal{D}/5.3)(\Gamma/5.3)$~cm for flare 1 and $r \lesssim 1.5 \times 10^{15}(\mathcal{D}/5.3)(\Gamma/5.3)$~cm for flare 2. 

\subsection{Spine-sheath geometry}
Outbursts observed in 3C~120 by {\it Fermi}/LAT are extreme both in terms of temporal and spectral characteristics. Both flares happened at very short timescales, of the order of hours, which is unusual as no such rapid outbursts have been observed before in $\gamma$-rays for any source of this kind. Intrestingly, there are no reports on increased activity for other wavelengths. For flare~1 several follow-up observations were reported: in optical \citep{ATel6542}, in X-rays \citep{ATel6606} and in near infrared \citep{ATel6613} however none of them reported any unusual source behaviour. All of them were carried out days after flaring activity hence it is difficult to directly connect those observations to $\gamma$-ray brightening. No significant change in flux has been observed in soft X-rays with the MAXI telescope\footnote{\texttt{http://maxi.riken.jp/top/index.php?cid=1\&jname=J0433+053}} \citep{Maxi} and in hard X-rays with the BAT instrument on board the Swift satellite\footnote{\texttt{http://swift.gsfc.nasa.gov/results/transients/weak/\\3C120.lc.txt}} \citep{Swift05}, for both flaring periods. We have no data on flaring activity in other wavelengths coinciding with flares 1~and~2 periods.

3C~120 $\gamma$-ray flux increase on very short timescale, observed spectral hardening, no flaring behaviour in X-rays and no indication of rapid changes in other wavelengths constitute a challenge for interpretation of those facts in terms of usual disk+jet models \citep{Tan15}. No indication of increased activity except for $\gamma$-rays suggest that high energy component and radiation at other wavelengths may be produced at different locations. This observation also supports the idea that BLRGs broadband spectra consist of disk-related and jet-related components. However, no change in hard X-rays during $\gamma$-ray flaring sets certain limitations on radiation mechanism responsible for high energy component in relativistic jet~\ie~assuming one-zone model for production of radiation in the jet, $\gamma$-ray emission is constrained by the constant X-ray flux.

$\gamma$-ray flux during flare 1 was $\sim50$~times and during flare 2 $\sim150$ about times larger than the average flux for 3C~120 for all {\it Fermi}/LAT data. Total apparent luminosity from the jet is $L=(\mathcal{D}^3/\Gamma) \eta_{\mathrm{rad}} \eta_{\rm e} \eta_{\mathrm{diss}}  L_{\mathrm{jet}}$ where $\eta_{\mathrm{diss}}$ is the total jet power dissipation efficiency, $\eta_{\rm e}$ describes what part of dissipated energy is transferred to electrons, $\eta_{\mathrm{rad}}$ is the radiative efficiency and $L_{\mathrm{jet}}$ is a total jet power before dissipation. Increase of total apparent luminosity can be achieved by either increase in jet bulk Lorentz factor $\Gamma$, more efficient jet power dissipation/radiative energy dissipation or increase in the total jet power. Let us focus on possible implications of either of these possibilities by assuming that the low-energy jet spectral component is synchrotron radiation whereas high-energy jet spectral component is external radiation Compton emission (ERC) (see section~\ref{sec:emission_mechanism}). The ERC luminosity can be estimated as $L_{\rm ERC} \propto \eta L_{\rm jet} \Gamma^2$ by its dependence on the jet power (where $\eta \equiv \eta_{\mathrm{rad}} \eta_{\rm e} \eta_{\mathrm{diss}}$) but also as $L_{\rm ERC} \propto u'_{\rm ext} \propto L_{\rm disk}$ where $u'_{\rm ext}$ is the external radiation energy density in the jet co moving frame which is a function of the accretion disk luminosity $L_{\rm disk}$. Synchrotron radiation luminosity can be estimated in a similar manner as $L_{\rm syn} \propto u'_{\rm B} \propto L_{\rm B} \propto \sigma L_{\rm jet}/(1+\sigma)$ where $u'_{\rm B}$ is the magnetic field energy density in the jet frame, $L_{\rm B}$ is the magnetic field energy flux and $\sigma$ is the jet magnetisation defined as $\sigma = L_{\rm B}/L_{\rm kin}$ where $L_{\rm kin}$ is the jet kinetic energy assumed to be dominated by cold protons. In a one-zone leptonic model $L_{\rm ERC}/L_{\rm syn} \propto u_{\rm ext} \Gamma^2/u'_{\rm B}$ thus we have
\[
L_{\rm syn} \propto \left(\frac{\sigma}{1+\sigma}\right) \left(\frac{L_{\rm jet}^2}{L_{\rm disk}}\right) \eta.
\]
If the increase in $\gamma$-ray band $\Delta L_{\rm ERC}$ was caused by the increase in the jet power $\Delta L_{\rm jet}$ this would lead to even more significant increase in synchrotron luminosity~\ie~$\Delta L_{\rm syn} \propto (\Delta L_{\rm jet})^2$. Increasing overall energy dissipation efficiency~\ie~$\Delta L_{\rm ERC} \propto \Delta \eta$ results in $\Delta L_{\rm syn} \propto \eta$. Finally, as the expression above is not a function of $\Gamma$, an increase of the $\Gamma$ factor causes significantly larger $\gamma$-ray flux [$\Delta L_{\rm ERC} \propto (\Delta \Gamma)^2$] while it does not affect synchrotron luminosity. Such simple analysis proves that first two possibilities have to be excluded as they would lead to increase in not only $\gamma$-rays but also other wavelengths which is not observed in case of 3C~120 flaring states. Third possibility can also be discarded because increase in $\Gamma$ factor would lead to narrower Doppler cone lowering total apparent luminosity at all wavelengths. A straightforward conclusion is that to explain such ``orphan'' flares in 3C~120 there is a need for a separate emission zone.
Such a zone should satisfy the following conditions:
\begin{itemize}
\item to have the ability to occasionally produce large $\gamma$-ray fluxes, strongly beamed towards the observer,
\item to have very large $\Gamma$ Lorentz factor to boost ERC radiation avoiding too large synchrotron luminosity,
\item be compact enough to address the very fast variability during flares.
\end{itemize}
We propose that such a component might be represented by the fast moving and wiggling spine.
Spine-sheath/layer jet structures have been considered in the literature in variety of models proposed to explain specific properties of jetted objects (see~\eg~\citealp{Cel01}; \citealp{Ghi05}; \citealp{Tav08}; \citealp{DArc09}; \citealp{Mim15}). In the case of a magnetically-arrested disk (MAD) scenario of AGN jets launching, which is likely to apply to 3C~120 \citep{Loh13}, faster spine and slower layer can naturally result from non-uniform mass loading of a jet at its base \citep{McK12}. In such a case mass loading is driven by interchange instabilities. Because these instabilities are non-axisymmetric, they may also be responsible for wiggling of a jet and its velocity variations. Jet direction can be also altered by the current-driven instabilities (\citealp{Nal12} and refs. therein). 
In this work we follow the spine-layer idea to model the spectra of flares 1 and 2 in 3C~120.


We assume that the conical jet with opening angle $\theta_{\mathrm{jet}}$ is stratified in such a way that it is composed of two elements: a fast conical spine with bulk Lorentz factor $\Gamma^{\rm{s}}$, opening angle $\theta^{\rm{s}}_{\mathrm{jet}}$ and a conical layer (``sheath'') with bulk Lorentz factor $\Gamma^{\rm{l}}=\Gamma$ and opening angle $\theta^{\rm{l}}_{\mathrm{jet}} = \theta_{\mathrm{jet}}$. The layer forms the main jet body being aligned with its axis and the spine forms an addition inside the jet. We assume that $\Gamma^{\rm{s}} \gg \Gamma^{\rm{l}}$ and $\theta^{\rm{s}}_{\mathrm{jet}} < \theta^{\rm{l}}_{\mathrm{jet}}$. We also assume that, while layer stays straight, the spine is subjected to change its direction with respect to jet axis and, therefore, with respect to the observer. In other words we assume layer's Doppler factor $\mathcal{D}^{\rm{l}}$ is constant while spine's Doppler factor $\mathcal{D}^{\rm{s}}$ is variable due to variable angle towards the observer $\theta^{\rm{s}}_{\rm{obs}}$. We associate this variability timescale with observer $\gamma$-ray variability.

%

Note that we do not discuss the origin of spine-layer configuration nor do we calculate spine-layer geometry from basic principles~\eg~using jet formation theories. However, since we assume that the layer actually forms the underlying and stable base of the jet its parameters, $\theta^{\rm{l}}_{\mathrm{jet}}$ and $\Gamma^{\rm{l}}$ can be inferred directly from observations. Spine's parameters on the other hand are model free parameters~\ie~they are adjusted to model source spectra.

Figure~\ref{fig:scheme} presents a schematic view of the geometry of the presented model.

\setcounter{figure}{5}
\begin{figure*}
  \includegraphics[width=.45\textwidth]{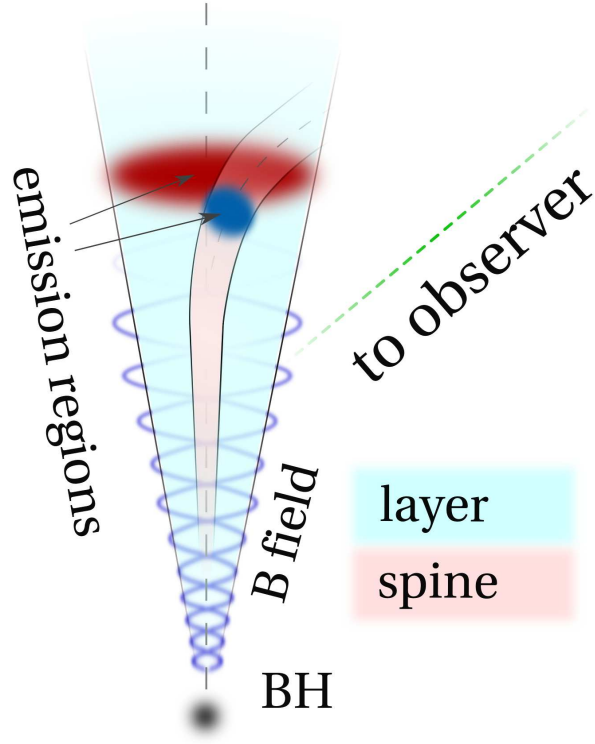}
 \caption{Schematic view of the adopted jet geometry of the spine-layer model.}
  \label{fig:scheme} 
\end{figure*}

\subsection{\boldmath{$\gamma$}-ray emission mechanisms}
\label{sec:emission_mechanism}
While the low energy jet radiation component is likely synchrotron emission the radiation mechanism for $\gamma$-rays remains unclear. Possible mechanisms include self-synchrotron Compton (SSC; \citealp{Mar92}; \citealp{Bloom96}) or external radiation Compton (ERC; \citealp{Der92}; \citealp{Sik94}) on broad-line region (BLR) and hot dust region (HDR) seed photons.

\citet{Pozo14} calculated the location of the thin-disk BLR in 3C~120 between 22 and 28 light days (l.d.) from the central BH with reverberation mapping technique. Similar results very obtained by \citet{Koll14} who found BLR to be stratified with He II emission line located at $12 \pm 7$~l.d. and H$\alpha$ line much further at $28.5\pm8.5$~l.d. away from the BH. For the purpose of this work we will adopt a mean BLR radius value of 25~l.d. which corresponds to  $r_{\rm{BLR}} = 6.5\times10^{16}$~cm~$\approx 0.02$~pc. Recent models of duty torus and BLR in AGN \citep{Cze11} and IR torus observations \citep{Kish11} suggest that the HDR characteristic radius $r_{\rm{HDR}} \sim 10 r_{\rm{BLR}}$ thus we set $r_{\rm{HDR}} = 6.5\times10^{17}$~cm~$\approx 0.2$~pc. External photon fields energy density in the jet co-moving frame for $r$ larger than the characteristic radius is $u'_{\rm{ext}} = \xi \xi_{\rm{CF}} \Gamma^2 L_{\rm{disk}} / 4 \pi r^2 c$ where $\xi_{\rm{CF}}$ is BLR (HDR) covering factor (we assume $\xi^{\rm{BLR}}_{\rm{CF}} = 0.1$ \citep{Sik09} and $\xi^{\rm{HDR}}_{\rm{CF}} = 0.3$ due to larger torus size) and $\xi \approx 0.1$ is a factor accounting for flat geometry of photon emission regions \citep{Jan15}. 3C~120 accretion disk flux at $\sim10$~eV is $\sim 1.5 \times 10^{-10}\,\rm{erg\,cm}^2\,{s}^{-1}$ resulting in disk luminosity $L_{\rm{disk}} \approx 1.7 \times 10^{45}~\rm{erg\,s}^{-1}$ after applying bolometric correction of $\sim 4.5$ from \citet{Rich06}. Similar value of disk luminosity was obtained by \citet{Ogle05}. We note that this value and our assumption concerning dusty torus geometry is consistent with observation that the inner edge of dusty torus in AGN is located approximately at graphite sublimation radius $r_{\rm{sub}} = 1.6 \times 10^{-5} L_{\rm{disk}}^{1/2} \approx 6.6 \times 10^{17}$~cm \citep{Mor12}. 

To find an efficient and dominant mechanism for $\gamma$-ray production in case of all data averaged spectra let us estimate the ratio between ERC and SSC luminosities produced in a uniform jet. For aforementioned values we obtain 
\[
u'_{\rm{ext}} \sim 3 \times 10^{-3} \left(\frac{\ds \xi^{\rm{ext}}_{\rm{CF}}}{\ds 0.1}\right)\left(\frac{\ds \Gamma}{\ds 5.3}\right)^2 \left(\frac{\ds r}{\ds r_{\rm{HDR}}}\right)^{-2}~\rm{erg~cm}^{-3}.
\]
Synchrotron radiation energy density in the jet frame is $u'_{{\rm syn}} = L_{{\rm syn}}/4 \pi \mathcal{D}^4 \theta_{\rm jet}^2 r^2 c$ \citep{Jan15} - with synchrotron bolometric luminosity $L_{\rm{syn}} \approx 1.6 \times 10^{44}~\rm{erg~s}^{-1}$ (see section~\ref{sec:model_fit} and Fig.~\ref{fig:base_model}) we get 
\[
\begin{array}{lcl}
u'_{{\rm syn}} & \sim & 1.5 \times 10^{-4} \left(\dfrac{\ds \mathcal{D}}{\ds 3.7}\right)^{-4} \left(\dfrac{\ds \theta_{\rm jet}}{\ds 0.19}\right)^{-2} \\
& & \times \left(\dfrac{\ds r}{\ds r_{\rm{HDR}}}\right)^{-2}\rm{erg~cm}^{-3}.
\end{array}
\]
The ratio between ERC and SSC luminosity is approximately 
\[
\begin{array}{lcl}
\dfrac{\ds L_{\rm{ERC}}}{\ds L_{\rm{SSC}}} & = & \left(\dfrac{\ds \mathcal{D}}{\ds \Gamma}\right)^2 \dfrac{\ds u'_{{\rm ext}}}{\ds u'_{{\rm syn}}} \\
                      & \approx & 10~\left(\dfrac{\ds \xi^{\rm{ext}}_{\rm{CF}}}{\ds 0.1}\right) \left(\dfrac{\ds \mathcal{D}}{\ds 3.7}\right)^{6} \left(\theta_{\rm jet}/0.19\right)^{2},
\end{array}
\] therefore ERC mechanism is dominating high energy radiative output. However, we note that this result strongly depends on Doppler factor $\mathcal{D}$ and jet opening angle $\theta_{\rm jet}$ - both these factor are poorly determined. Lowering the Doppler factor to a value of $2.4$ (see section~\ref{sec:location}) or lowering $\theta_{\rm jet}$ to a value of $\sim 0.02$ may result in indication that SSC mechanism is favoured [see \citet{Tan15}]. Thus, despite our estimation we conclude that both ERC and SSC mechanisms can be equally viable to explain the averaged data for 3C~120.

Our conclusion has to be altered in the case of observed $\gamma$-ray flares where we use the sum of the spine and layer radiation to model source spectra. In such case production of $\gamma$-ray emission in spine via SSC mechanism is very unlikely as it would impose significant flux increase in radio (due to synchrotron emission) and X-ray band (due to spectral broadness of SSC component). ERC mechanism on external photons from BLR(HDR) is therefore favoured as it can account for increased $\gamma$-ray emission without the increase of flux in X-rays (see~3.2) provided its production in separate component (spine).

In the spine-layer geometry model, during the period of source quiescence the total radiative output is dominated by the layer emission. Indeed, relatively low $\Gamma$ factor and large viewing angles cause the jet emission to be significantly lower than the accretion-related emission. At this period of time the inner, fast spine may or may not exist - this fact is not resembled in the source spectra due to even larger viewing angles and spine power being lower than power carried by the layer (see 3.2). However, whenever the spine bends towards the observer (see Fig.~\ref{fig:scheme} and section 3.2) the spine emission starts to dominate the total radiative output. In terms of spectral characteristics the ratio between ERC (from spine) and synchrotron (from layer) emission is $L^{\rm s}_{\rm{ERC}}/L^{\rm l}_{\rm{syn}} \propto u'^{\rm s}_{\rm{ext}}/u'^{\rm l}_{\rm{B}} \propto {\Gamma^{\rm s}}^2 \times u_{\rm{ext}}/u'^{\rm l}_{\rm{B}}$~\ie~it is largely dominated by spine ERC radiation due to its much larger $\Gamma^{\rm s}$ factor.

Our estimates presented in previous paragraph (assuming values of Doppler factor observed for 3C~120) locate the blazar zone at sub-parsec scales deep inside BLR. In fact, as shown by \citet{Jan15} at distances lower than $5 \times 10^{16}$~cm from the central BH jet environment is dominated by photons from the accretion disk. In such circumstances $\gamma$-ray absorption via pair production is inevitable already at GeV energies \citep{Pou10}. The bent spine scenario, however, may provide a much larger size of the estimated emission region because of larger values of both Doppler factor $\mathcal{D}$ and Lorentz factor $\Gamma$ (see 5.1)~\ie~the estimated size $R$ can be increased maximally by a factor $(\Gamma^{\rm s}/\Gamma^{\rm l})^2$ (assuming $\theta_{\rm{obs}} = 1/\Gamma^{\rm{s}}$) with respect to the value obtained with the assumption that the emission region is calculated with full conical jet scenario.

The spine-layer geometry essentially introduces a second emission zone and, therefore, inevitable interaction between radiation produced in both spine and layer regions. The most significant radiative component to take into account is layer's (spine's) synchrotron radiation Compton up-scattered in the spine (layer) region. However, due to geometry synchrotron radiation produced in spine would be significantly diluted in the much larger layer emission region \citep{Ghi05}. Therefore, it is most important to estimate the spine Compton radiation on layer's synchrotron photons. The ratio between ERC emission in spine $L^{\rm s}_{\rm ERC}$ and external synchrotron Compton emission in spine $L^{\rm s}_{\rm ESC}$ is
\[
\frac{L^{\rm s}_{\rm ERC}}{L^{\rm s}_{\rm ESC}} \propto \frac{u'^{\rm s}_{\rm ext}}{u'^{\rm s}_{\rm syn, l}} \propto \frac{{\Gamma^{\rm s}}^2 u_{\rm ext}}{{\Gamma^{\rm sl}}^2 u'^{\rm l}_{\rm syn}},
\]
where $u'^{\rm s}_{\rm syn, l}$ denoted layer's synchrotron radiation energy density in spine and $\Gamma^{\rm sl} = \Gamma^{\rm s} \Gamma^{\rm l} (1-\beta^{\rm s} \beta^{\rm l})$ is an effective Lorentz factor between spine and layer \citep{Ghi05}. It can be shown that $\Gamma^{\rm sl} \approx \Gamma^{\rm s}/2\Gamma^{\rm l}$ for $\Gamma^{\rm s} \gg \Gamma^{\rm l}$. Using values calculated before we get $L^{\rm s}_{\rm ERC}/L^{\rm s}_{\rm ESC} \sim 70$ hence spine Compton up-scattered synchrotron radiation from layer is much less luminous as compared to spine's ERC radiation and may be omitted in further considerations.

\subsection{Model parameters and spectral modelling}
\label{sec:model_fit}
For spectral modelling we use numerical code already described in \citet{Jan15}. We point out main assumptions of the numerical model below.

Jet energy dissipation takes place in an active region of conical jet, at distances between $[r,2r]$ from the central BH and production of radiation is followed further up to $20r$. Electron evolution and production of radiation is calculated in a steady-state manner \citep{Sik13}. 

We follow the evolution of electron energy distribution in the region of interest by solving the kinetic equation for relativistic electrons \citep{Mod03} which can be presented in the form
\[
 \frac{\p N_{\gamma} (r)}{\p r} = - \frac{\p}{\p \gamma} \left( N_{\gamma}(r) \frac{{\rm d} \gamma}{{\rm d} r} \right) + \frac{ Q_{\gamma}(r) }{ c \beta \Gamma } \, ,
 \]
where $N_{\gamma}$ is the number of electrons per energy bin, $\beta = \sqrt{\Gamma^2-1}/\Gamma$, 
${\rm d}\gamma/{\rm d}r= ({\rm d}\gamma/{\rm d}t')/(\beta c \Gamma)$, ${\rm d}\gamma/{\rm d}t'$ are the electron 
energy loss rates as measured in the jet co-moving frame, and $Q_{\gamma}(r)$ is the broken power-law electron injection function defined within $[\gamma_{\rm{min}},\gamma_{\rm{max}}]$ energy range, with spectral indices $p_{1}$~and~$p_{2}$. Injection function normalisation and electron spectrum break energy $\gamma_{\rm{b}}$ are calculated based on several parameters~\ie~jet magnetisation $\sigma$, jet pair content $n_{\rm e}/n_{\rm p}$, jet production efficiency $\eta_{\rm j}$, fraction of energy transferred to electrons $\eta_{\rm e}$, jet magnetic field $\rm{B}$ and energy dissipation efficiency $\eta_{\rm diss}$. These parameters also allow to calculate jet energetics~\ie~total jet power $L_{\rm jet}$, protons kinetic energy $L_{\rm P}$ and magnetic energy flux $L_{\rm B}$ (see \citealp{Jan15} for details).

Synchrotron, SSC and adiabatic electron energy loss rates are calculated using the procedure presented by \citet{Mod03}. ERC energy loss rates and luminosities of all radiation mechanisms as well the description of the adopted here planar model of external radiation sources BLR and HDR were described in our previous works (\citealp{Sik13}, \citealp{Jan15}).

For all {\it Fermi}/LAT data we set the location of the blazar zone at $r=2.6 \times 10^{17}$~cm (see 3.1) assuming $\Gamma=5.3$ and $\theta_{\rm{jet}} = 1/\Gamma$. We assume the central BH mass $M_{\rm{BH}} = 4.6 \times 10^{8} M_{\sun}$ \citep{Pozo14} which for accretion disk luminosity $L_{\rm{disk}} = 1.7 \times 10^{45}~\rm{erg s}^{-1}$ and assumed accretion disk radiative efficiency $\eta_{\rm{disk}}=0.1$ results in accretion power $\dot{M}c^2 \sim 1.7 \times 10^{46}~\rm{erg\,s}^{-1}$ and Eddington radio of $\sim 0.03$. We note that BH mass estimation is very uncertain in 3C~120 as many works reported much lower values: $5.5 \times 10^{7} M_{\sun}$ \citep{Pet04}, $3.0 \times 10^{7} M_{\sun}$ \citep{Mar09} or $5.7 \times 10^{7} M_{\sun}$ \citep{Pozo12}. To model accretion-related emission in 3C~120 we use a radio-loud quasar radiation template \citep{Sha11} with jet radio emission truncated at $\sim 10^{13}$~Hz. We match the template with the X-ray data in $1-6$~keV energy band \citep{ATel6606}.

To estimate other input parameters of our jet emission model we choose them so that the calculated spectra match the observations. Having set an average value of the angle towards the observer $\theta_{\rm{obs}} = 0.26 \sim 15\degr$ normalisation of calculated emission depends on total jet radiative efficiency, fraction of jet power effectively channelled to electrons in dissipation region and initial power carried by jet itself. Radiative efficiency at sub-parsec scales~\ie~inside dense photon fields from BLR and HDR was estimated to be $>0.5$ \citep{Jan15} and initial jet power cannot significantly exceed accretion power. As 3C~120 is an FR~I type of radio source energy dissipation $\eta_{\rm{diss}}$ cannot be too low. The upper limit on $\eta_{\rm{diss}}$ can be inferred from the fact that the spectral peak $\nu_{\rm{peak}}$ in $\gamma$-rays is located at energies lower than $100$~MeV (see Fig.~\ref{fig:base_spectra}). For fixed values of injected electrons spectral indices the break in broken power law spectrum $\gamma_{\rm{break}} \propto \eta_{\rm{diss}}$. Since $\nu_{\rm{peak}} \sim \mathcal{D}^2 \gamma_{\rm{break}}^2 \nu_{\rm{ext}}$ we find that the upper limit on $\gamma_{\rm{break}}$ is $\sim 10^{3}$ (assuming ERC process on BLR photons with $\nu_{\rm{ext}} = \nu_{\rm{BLR}} = 10~\rm{eV}$). We assume that energy dissipation efficiency is $\eta_{\rm{diss}} = 0.3$ (which results in $\gamma_{\rm break} \approx 6 \times 10^{2}$) and half of that energy is transferred to electrons.

For injected electron energy spectrum we set spectral indices to $p_{1}=0.5$ for $\gamma \le \gamma_{\rm break}$ and $p_{2}=2.4$ for $\gamma > \gamma_{\rm break}$. We set the minimum electron energy to $\gamma_{\rm{min}} = 1.0$ and the maximum electron energy to $\gamma_{\rm{\max}}=2\times10^{4}$ which results in high energy tail of jet emission to match the {\it Fermi}/LAT spectrum. We note that the choice of $\gamma_{\rm{\max}}$ is very dependent on preferred $\gamma$-ray emission mechanism~\ie~if SSC was the dominant component at MeV-GeV energies reproducing {\it Fermi}/LAT spectra would require much larger value of $\gamma_{\rm{\max}} \sim 10^{6}$ \citep{Tan15}.

We find that choosing jet power $L_{\rm{jet}} \sim 1.3 \times 10^{45}~\rm{erg\,s}^{-1}$ is sufficient to match calculated spectra with the data. Obtained value of jet power is smaller than accretion power by a factor of $\sim 10$. This is in contrast with result by \citet{Tan15} that $L_{\rm{jet}} > L_{\rm{acc}}$. However, we note that for our calculations we used much higher value of accretion disk radiative power as well as our model favours ERC as primary source of $\gamma$-ray emission. 

The choice of the magnetic field strength is dictated by the normalisation of radio data. We set the magnetic field $B'= 0.2~\rm {G}$ at $0.1$~pc from the central BH. Such value results in magnetic field energy flux $L_{\rm{B}} \sim 4.5 \times 10^{43}~\rm{erg\,s}^{-1}$. Assumed model parameters lead to kinetic energy of cold protons $L_{\rm P} \sim 9 \times 10^{44}~\rm{erg\,s}^{-1}$ therefore jet magnetisation parameter $\sigma = L_{\rm B}/L_{\rm P} = 0.05$. This result indicate that in the emission region at sub-parsec scales relativistic jet in 3C~120 is already dominated by kinetic flux~\ie~transition from initially Poynting dominated jet to matter dominated flow must have occurred closer to the central BH. 

Figure~\ref{fig:base_model} presents broadband 3C~120 spectra for all {\it Fermi}/LAT data calculated with described numerical model.

Since modelling ``orphan'' flares in 3C~120 requires an additional spine component we assume that the layer forms the underlying steady part of the jet in such a way that the total jet radiation consists of layer emission with additional spine emission. Note however, that during quiescence period the spine may also exist in the jet yet due to large $\Gamma^{\rm s}$ and large viewing angles its radiation is Doppler de-boosted~\ie~even though the spine carries significant part of the total jet power it remains ``invisible''. However, sudden change of direction of the spine towards the observer results in sudden increase of the Doppler factor $\mathcal{D}^{\rm s}$ so that the spine contribution to total jet radiation becomes significant. Simplicity of such model implies only few additional parameters to be determined to match observed spectra, namely: spine Lorentz factor $\Gamma^{\rm s}$, spine opening angle $\theta^{\rm s}$ and an angle towards the observer $\theta^{\rm s}_{\rm obs}$. We do not make any further assumptions concerning the magnetic field in spine except for keeping spine magnetisation identical to magnetisation of the layer. This attitude is dictated by model simplicity~\ie~except for the spine's magnetic field substantially higher than layer's, total radiative output is dominated by ERC emission due to much higher Lorentz factor thus synchrotron and SSC components are greatly weakened in spine emission. 

We choose aforementioned spine parameters so that the calculated spectra matches the {\it Fermi}/LAT data during outbursts 1~and~2. For flare 1 we choose $\Gamma^{\rm s, 1} = 20$ and for flare 2 we choose $\Gamma^{\rm s, 2} = 40$. We choose spine opening angles so that they follow a relation $\theta^{\rm s} = 0.8/\Gamma^{\rm s}$~\ie~slightly narrower with respect to the Lorentz factor than assumed layer opening angle. Such choice results in $\theta^{\rm s, 1} = 0.04$ and $\theta^{\rm s, 2} = 0.02$ during flares 1 and 2, respectively. We also assume that the spine is bent towards the observer so that it is seen exactly on the border of the Doppler cone~\ie~$\theta^{\rm s, 1,2}_{\rm obs} = \theta^{\rm s, 1,2}$. Resulting Doppler factors are $\mathcal{D}^{\rm s, 1} \approx 25$ and $\mathcal{D}^{\rm s, 2} \approx 50$. Such choice of $\Gamma^{\rm s}$ and $\mathcal{D}^{\rm s}$ parameters locates the emission region in the spine at $r^{1,2} \approx 1.4 \times 10^{17}~\rm{cm}~\approx 0.04~\rm{pc}$ -- this value is almost identical for both flares despite significant difference in $\gamma$-ray flux and assumed spine Lorentz factor. Estimated location of the blazar zone in the spine is just outside BLR thus it enables avoiding significant absorption due to $\gamma\gamma$ pair production. With such parameters we are able to reproduce source spectra during flaring states. Results of spectral energy distribution (SED) modelling are presented in Fig.~\ref{fig:flares_model}.

Calculated spine powers are much lower than power carried by the layer. In case of flare 1 the ratio between spine and layer power $L_{\rm s}/L_{\rm l} \approx 0.14$ and $\approx 0.10$ for flare 2~\ie~the layer carries majority of the jet power. 

\setcounter{figure}{6}
\begin{figure*}
  \includegraphics[width=.75\textwidth]{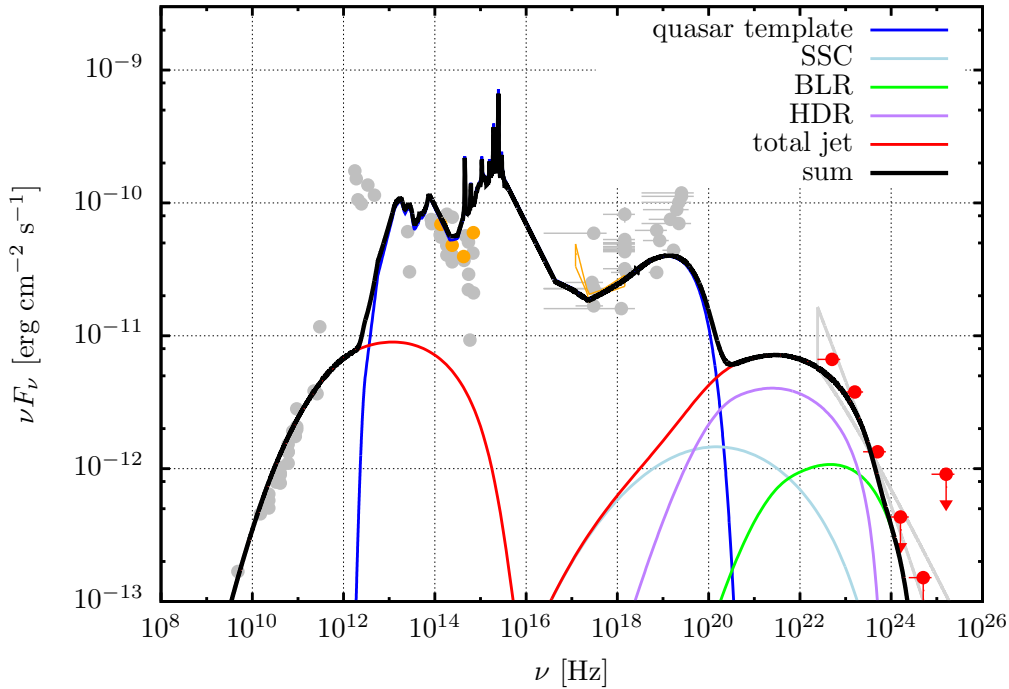}
  \caption{3C~120 broadband spectrum; red points are {\it Fermi}/LAT data points for all {\it Fermi}/LAT data (from August~4th, 2008 to May~26th, 2015; see section 2) where arrows indicate $95\%$~upper limits and light-grey ``butterfly'' plot presents best-fit power law $\gamma$-ray spectra. Orange points indicate optical and X-ray data from August 2014 (see section 3.2) and grey points are archived data collected from NED database. Solid blue line presents radio-loud quasar radiation template taken from \citet{Sha11}. BLR and HDR stand for ERC radiation on BLR and HDR seed photons, respectively. Red solid line corresponds to total jet radiative output and black solid line is a sum of accretion-related and jet radiation.}
  \label{fig:base_model}
\end{figure*}

\setcounter{figure}{7}
\begin{figure*}
  \includegraphics[width=.45\textwidth]{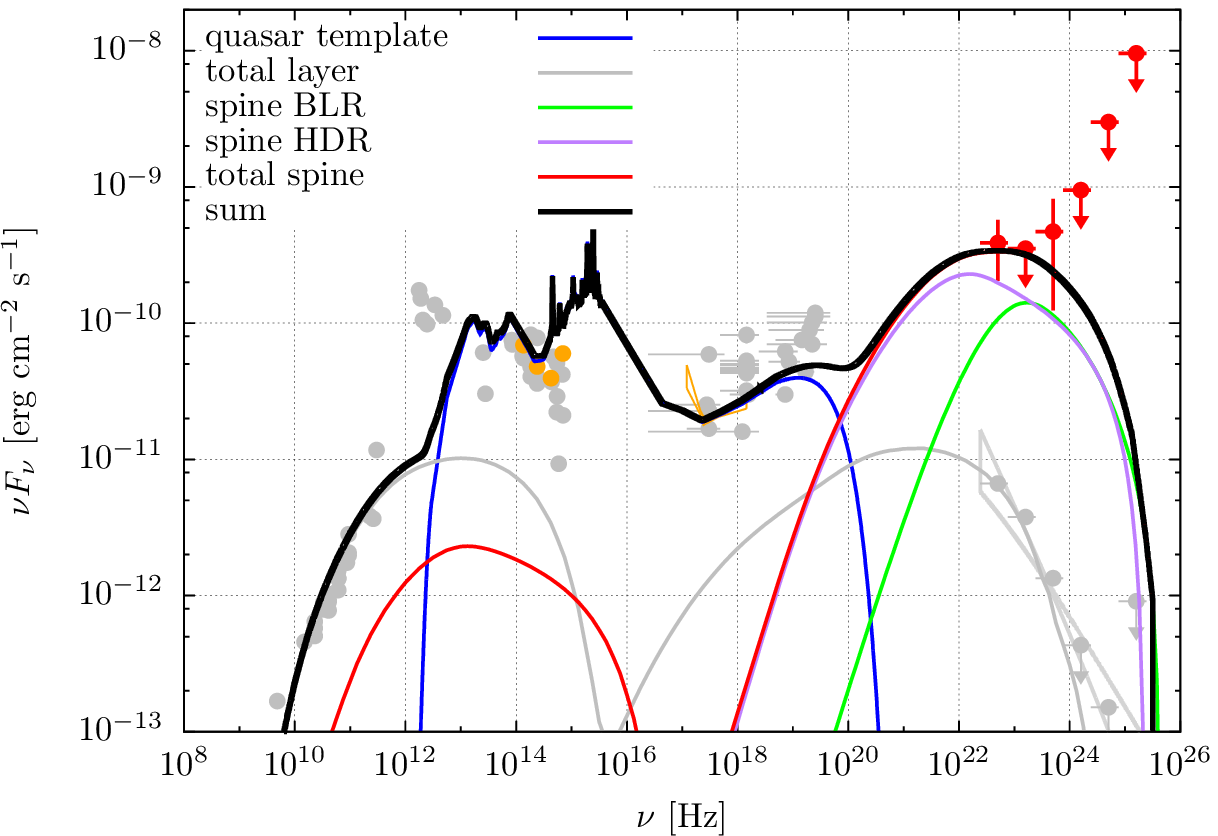}
  \includegraphics[width=.45\textwidth]{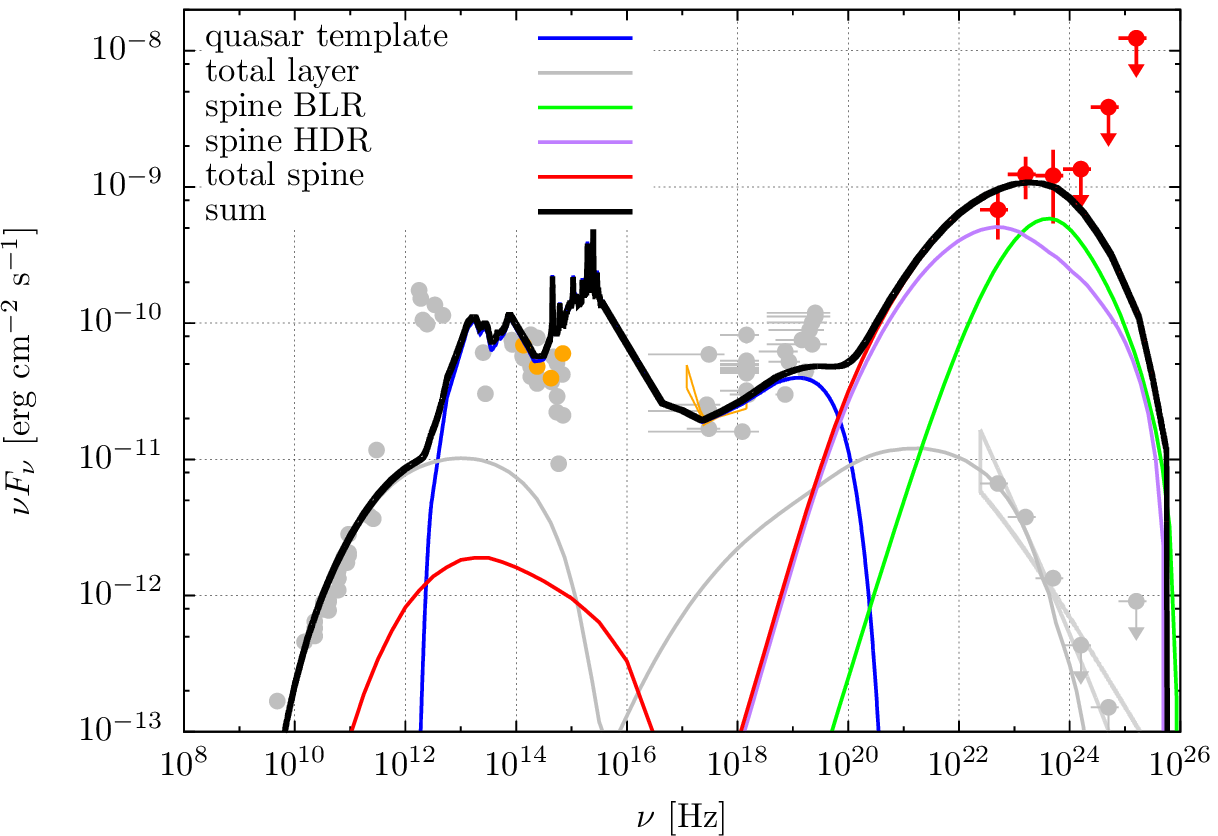}
  \caption{3C~120 broadband spectrum for flare 1 (\emph{left panel}) and flare 2 (\emph{right panel}); red points are {\it Fermi}/LAT data points for flares 1~and~2 where arrows indicate $95\%$~upper limits and light-grey ``butterfly'' plot presents best-fit power law $\gamma$-ray spectra for all {\it Fermi}/LAT data. Orange points indicate optical and X-ray data from August 2014 (see section 3.2) and grey points are archived data collected from NED database. Solid blue line presents radio-loud quasar radiation template taken from \citet{Sha11}. BLR and HDR stand for ERC radiation on BLR and HDR seed photons, respectively. Gray solid line presents layer radiation, red solid line corresponds to total spine radiative output and black solid line is a sum of accretion-related and jet (spine+layer) radiation.}
  \label{fig:flares_model}
\end{figure*}

\setcounter{table}{1}
\begin{table*}
\caption{Parameters used in numerical simulations. Single, centred values are common for all {\it Fermi}/LAT and flare 1 \& 2 data.}
\begin{center}
\begin{tabular}{|lccc|}
\hline
\hline
Parameter & All data & Flare 1 & Flare 2 \\
 & (layer) & \multicolumn{2}{c}{(spine)} \\
\hline
Black hole mass $M_{\rm BH}$  & \multicolumn{3}{c}{$4.6 \times 10^{8} M_{\sun}$} \\ 
Accretion rate $\dot{M}$  & \multicolumn{3}{c}{$0.3~L_{\rm Edd}/c^2$} \\ 
Accretion disk radiative efficiency $\eta_{\rm disk}$  & \multicolumn{3}{c}{$0.1$} \\
Energy dissipation efficiency $\eta_{\rm diss}$  & \multicolumn{3}{c}{$0.3$} \\
Jet Lorentz factor $\Gamma$  & $5.3$  & $20.0$  & $40.0$  \\
Doppler boosting factor $\mathcal{D}$ & $3.7$  & $25$  & $50$  \\
Fraction of energy transferred to electrons $\eta_{\rm e}$ & \multicolumn{3}{c}{$0.5$} \\
Magnetic field $\rm{B}$ at $0.1$~pc [G]& $0.2$ & $0.1$ & $0.25$ \\
Jet magnetisation $\sigma$ & \multicolumn{3}{c}{$0.05$} \\
Pair content $n_{\rm e}/n_{\rm p}$  & \multicolumn{3}{c}{$0.5$} \\
Electron injection function indices $p_{1}, p_{2}$  & \multicolumn{3}{c}{$0.5, 2.4$} \\
Min. and max. injection energies $\gamma_{\rm min}, \gamma_{\rm max}$  & \multicolumn{3}{c}{$1, 2 \times 10^{4}$} \\
Jet opening angle $\theta_{\rm j}$  & $0.19$ & $0.04$ & $0.02$ \\
                                  & $1/\Gamma$ & $1/(0.8\Gamma)$ & $1/(0.8\Gamma)$ \\
Observing angle $\theta_{\rm obs}$  & $0.26$ & $0.04$ & $0.02$ \\
Location of emission region $r$ & $2.6 \times 10^{17}$~cm & $9.5 \times 10^{16}$~cm & $7.3 \times 10^{16}$~cm \\
BLR photons energy $\nu_{\rm{BLR}}$ & \multicolumn{3}{c}{$10\,{\rm eV}$} \\
HDR photons energy $\nu_{\rm{HDR}}$ & \multicolumn{3}{c}{$0.06 - 0.6 \, {\rm eV}$} \\
BLR radius $r_{\rm{BLR}}$ & \multicolumn{3}{c}{$6.5 \times 10^{16}$~cm~$=0.02$~pc} \\
HDR radius $r_{\rm{HDR}}$ & \multicolumn{3}{c}{$6.5 \times 10^{17}$~cm~$=0.2$~pc} \\
BLR covering factor $\xi_{\rm BLR}$   & \multicolumn{3}{c}{$0.1$} \\
HDR covering factor $\xi_{\rm HDR}$  & \multicolumn{3}{c}{$0.3$} \\
\hline
Accretion disk luminosity $L_{\rm{disk}}~[\rm{erg~s}^{-1}]$ & \multicolumn{3}{c}{$1.7 \times 10^{45}$} \\
Jet power $L_{\rm{jet}}~[\rm{erg~s}^{-1}]$ & $1.3 \times 10^{45}$ & $1.8 \times 10^{44}$ & $1.3 \times 10^{44}$ \\
Kinetic energy of protons $L_{\rm P}~[\rm{erg~s}^{-1}]$ & $9 \times 10^{44}$ & $1.2 \times 10^{44}$ & $9.0 \times 10^{43}$ \\
Magnetic energy flux $L_{\rm B}~[\rm{erg~s}^{-1}]$ & $4.5 \times 10^{43}$ & $6.0 \times 10^{42}$ & $4.5 \times 10^{42}$ \\
\hline
\hline
\end{tabular}
\end{center}
\label{tab:param}
\end{table*}

\section{Conclusions}
\label{sec:con}
{\it Fermi}/LAT data analysis suggests that although it is a rather faint $\gamma$-ray source it is worth monitoring especially as it occasionally shows very luminous flares. Detailed data analysis of two recently reported outbursts (see Fig.~\ref{fig:base_lc}) proved not only extraordinary $\gamma$-ray flux increase (by a factor of $\sim 50$ for flare 1 and $\sim 150$ in case of flare 2) but also very short timescales of this phenomenon. As we argue in section 3.2 it is rather impossible to explain these factors with a one-zone emission model. We demonstrated that an introduction of the fast, narrow and wiggling spine component is a model which meets all the required properties.

Such model proves to be very robust in explaining 3C~120 flaring behaviour as it can account for powerful $\gamma$-ray flux increases without significant flux changes in other wavelengths while keeping the emission region small. Model robustness is also strengthened by the fact that the introduction of second jet component does not lead to doubling of model parameters. Indeed, the spine-layer scenario introduces only two additional model parameters: the spine Lorentz factor $\Gamma^{\rm s}$ and its opening angle $\theta^{\rm s}$. However, these parameters are merely just spine flux normalisation constants and there is a need for a viable theoretical model that could put constraints on their values and relations towards other jet properties.

Positive detection for time bins as short as one hour (or even less in case of flare 2) must lead to extremely small emitting zone if interpreted as coherent emission from localised active region in a horizontally uniform jet. This observation inevitably localises the $\gamma$-ray active region very close to the central BH unless the jet is very fast and pointing towards the observer which leads to high Doppler factors and, therefore, emission region in the jet co moving frame is much larger.

We model the 3C~120 SED assuming its total radiation consists of the accretion-related and jet-related components. In case of all data, averaged {\it Fermi}/LAT spectrum can be explained as having pure SSC origin \citep{Tan15} or as being dominated by ERC radiation making use of strong BLR and HDR radiation in the jet environment. Note that neither of these models is favoured as the choice between the dominant $\gamma$-ray production mechanism is very dependent on the Doppler factor and observations indicate that this value is highly variable. \citet{Cas15} claims that the observed $\gamma$-ray flux depends not only on the energetics of emission region but also on the changing jet orientation with respect to the observer. This observation supports the idea of the ``wiggling'' jet. We note that, although such conclusion has been made by analysing VLBI radio data, the same conclusion may be true in the case of the fast, narrow spine. In general, both the layer and the spine may change their directions, but due to large difference in their Lorentz factors observational effects are much more prominent in the case of the spine.

We model 3C~120 flares 1~and~2 by varying spine's Lorentz factor and by changing its direction towards the observer. The increase of $\Gamma$ factor in spine by four times, in case of flare 1, and by eight times, in the case of flare 2, with respect to the average value of $5.3$ are enough to properly reproduce the observed $\gamma$-ray spectra. We note that this applies not only to flux normalisation  but also reproduces the $\gamma$-ray spectral shape. During flares 3C~120 $\gamma$-ray spectrum was much harder ($\Gamma \sim 2.0$) that the averaged value ($\Gamma \sim 2.7$). The characteristic energy of BLR photons ($\sim 10~\rm{eV}$) and HDR photons ($\sim 0.6~\rm{eV}$) together with calculated electron distribution spectral brake $\gamma_{\rm b}$ and shift towards higher energies by larger $\Gamma^{\rm s}$ localises the ERC(BLR+HDR) spectral peak in the right position matching the observations. We also point out that both in all data spectra and during flares 3C~120 is visible by {\it Fermi}/LAT only up to energies of about several GeVs. This may indicate that a part of the high energy radiation is absorbed via $\gamma\gamma$ pair production on dense BLR photon field.

The calculated powers carried by spine and layer in the form of kinetic energy of cold protons and Poynting flux indicate that majority of the total jet energy is contained in the layer. This is in contrast to models proposed by \citet{Ghi05} where most of the total jet power was carried by the spine exceeding layer's power by at least several times.

Spine Lorentz factors proposed by our modelling are much different from average value of $5.3$ so they are different from each other. We point out that the proposed model which assumes different $\Gamma^{\rm s}$ factors for flares 1~and~2 is not unique~\ie~it is possible to get similar results with different set of input parameters. It is especially important to notice that the observed radiation is very sensitive to assumed Doppler factors, hence, the direction of the fast spine towards the observer. It is very plausible that both flares could be modelled with identical spine Lorentz factor with different direction towards the observer as the main reason for different observed $\gamma$-ray fluxes. 

Another way of having such extreme values in a single source is that the spine-layer model might be just a coarse simplification and yet an indication of a much more complex jet structure. Instead of a sharp division into fast and slow components one may think of a smooth transition from slower, outer parts of the jet towards faster, inner regions~\eg~described by a Gaussian profile $\Gamma(R)$ where $R$ is the distance from the jet axis. Depending on the broadness of such Gaussian $\Gamma$ factors distribution observed radiation could be characterised by different $L_{\rm ERC}/L_{\rm syn}$ ratio. Also the variability of this ratio would be very sensitive on the jet $\Gamma(R)$ profile~\ie~small short-term $\gamma$-ray variability would indicate rather broad profile while rapid, powerful high energy flares could be explained with peaked, narrow distribution similar to model proposed in this work. However, detailed and quantitative model of such proposition is required to assess its validity.

\section*{Acknowledgements}
This research has made use of the MAXI data provided by RIKEN, JAXA and the MAXI team.
We acknowledge financial support by the Polish NCN grant DEC-2012/07/N/ST9/04242.


\begin{thebibliography}{99}

\bibitem[\protect\citeauthoryear{Abdo \ea}{2009}]{Abdo09} Abdo, A. A., Ackermann, M., Ajello, M., \ea~({\it Fermi}/LAT Collaboration) 2009, ApJ, 707, 55
\bibitem[\protect\citeauthoryear{Abdo \ea}{2010}]{Abdo10} Abdo, A. A., Ackermann, M., Ajello, M., \ea~({\it Fermi}/LAT Collaboration) 2010, ApJ, 719, 1433
\bibitem[\protect\citeauthoryear{Abdo \ea}{2010}]{Abdo10b} Abdo, A. A., Ackermann, M., Ajello, M., \ea~({\it Fermi}/LAT Collaboration) 2010, ApJ, 720, 912
\bibitem[\protect\citeauthoryear{Ackermann \ea}{2015}]{Ack15} Ackermann, M., Ajello, M., Atwood, W., \ea~({\it Fermi}/LAT Collaboration) 2015, arXiv:1501.06054
\bibitem[\protect\citeauthoryear{Barthelmy \ea}{2005}]{Swift05} Barthelmy, S. D., Barbier, L. M., Cummings, J. R.,~\ea~2005, Space Science Reviews, v120:143-164
\bibitem[\protect\citeauthoryear{Bloom \& Marscher}{1996}]{Bloom96} Bloom, S. D., \& Marscher, A. P., 1996, ApJ, 461, 657
\bibitem[\protect\citeauthoryear{Brown \& Adams}{2012}]{Bro12} Brown, A. M. \& Adams, J., 2012, MNRAS, 421, 2303-2309 
\bibitem[\protect\citeauthoryear{Casadio \ea}{2015}]{Cas15} Casadio, C., Gómez, J. L., Grandi, P., \ea~2015, arXiv:1505.03871
\bibitem[\protect\citeauthoryear{Cash}{1979}]{Cash79} Cash, W., 1979, ApJ, 228, 939-947
\bibitem[\protect\citeauthoryear{Celotti \ea}{2001}]{Cel01} Celotti, A., Ghisellini, G., \& Chiaberge, M., 2001, MNRAS, 321, L1-L5 
\bibitem[\protect\citeauthoryear{Czerny \& Hryniewicz}{2011}]{Cze11} Czerny, B., \& Hryniewicz, K., 2011, A\&A, 525, L8
\bibitem[\protect\citeauthoryear{Chatterjee \ea}{2009}]{Chatt09} Chatterjee, R., Marscher, A. P., Jorstad, S. G., \ea~2009, ApJ, 704, 1689-1703
\bibitem[\protect\citeauthoryear{D'Arcangelo \ea}{2009}]{DArc09} D'Arcangelo, F. D., Marscher, A. P., Jorstad, S. G., \ea~1009
\bibitem[\protect\citeauthoryear{Dermer \ea}{1992}]{Der92} Dermer, C. D., Schlickeiser, R., \& Mastichiadis, A., 1992, A\&A, 256, L27
\bibitem[\protect\citeauthoryear{Ghisellini \ea}{2005}]{Ghi05} Ghisellini, G., Tavecchio, F., \& Chiaberge, M., \ea~2005, A\&A, 432, 401-410
\bibitem[\protect\citeauthoryear{Grandi \& Palumbo}{2007}]{Gra07} Grandi, P., \& Palumbo,~G.~G.~C., ~2007, ApJ, 659, 235-240
\bibitem[\protect\citeauthoryear{{\it Fermi}/LAT Collaboration}{2015}]{FLatTeam} Fermi LAT Weekly Report N. 359
\bibitem[\protect\citeauthoryear{Hasan \ea}{2014}]{ATel6613} Hasan, I., MacPherson, E., Buxton, M., \ea~2014,  ATel \#6613
\bibitem[\protect\citeauthoryear{Hovatta \ea}{2009}]{Hov09} Hovatta, T., Valtaoja, E., Tornikoski, M., \ea~2009, A\&A, 494, 527-537
\bibitem[\protect\citeauthoryear{Janiak \ea}{2015}]{Jan15} Janiak, M., Sikora, M. \& Moderski, R.~2015, MNRAS, 449, 431-439
\bibitem[\protect\citeauthoryear{Jorstad \ea}{2005}]{Jor05a} Jorstad, S.~G., Marscher, A. P., Lister, M. L., \ea~2005, ApJ, 130, 1418-1465
\bibitem[\protect\citeauthoryear{Kataoka \ea}{2011}]{Kat11} Kataoka,~J., Stawarz, {\L}.; Takahashi, Y., \ea~2011, ApJ, 740, 29
\bibitem[\protect\citeauthoryear{Kishimoto \ea}{2011}]{Kish11} Kishimoto,~M., Hoenig, S. F.,, Robert Antonucci, R., \ea~2011, A\&A, 536, 78
\bibitem[\protect\citeauthoryear{Kollatschny \ea}{2014}]{Koll14} Kollatschny, W., Ulbrich, K., Zetzl, M., \ea~ 2014, A\&A, 566, A106
\bibitem[\protect\citeauthoryear{Lohfink \ea}{2013}]{Loh13} Lohfink, A. M., Reynolds, C. S., Jorstad, S. G., \ea~2013, ApJ, 772, 83
\bibitem[\protect\citeauthoryear{Lohfink \ea}{2014}]{ATel6606} Lohfink, A. King, A. L., Reynolds, C., \ea~2014,  ATel \#6606
\bibitem[\protect\citeauthoryear{Maraschi \ea}{1992}]{Mar92} Maraschi, L., Ghisellini, G., \& Celotti, A., 1992, ApJ, 397, L5
\bibitem[\protect\citeauthoryear{Marshall \ea}{2009}]{Mar09} Marshall, K., Ryle, W. T., Miller, H. R. \ea~2009, ApJ, 696, 601-607
\bibitem[\protect\citeauthoryear{Matsuoka \ea}{2009}]{Maxi} Matsuoka, M., Kawasaki, K., Ueno, S., \ea~2009, PASJ, 61, 999
\bibitem[\protect\citeauthoryear{McKinney \ea}{2012}]{McK12} McKinney, J. C., Tchekhovskoy, A. \& Blandford, R. D., 2012, MNRAS, 423, 3083-3117
\bibitem[\protect\citeauthoryear{Mimica \ea}{2015}]{Mim15} Mimica, P., Giannios, D., Metzger, B. D., \ea~2015, MNRAS, 450, 2824-2841
\bibitem[\protect\citeauthoryear{Moderski \ea}{2003}]{Mod03} Moderski, R., Sikora, M., \& B{\l}a{\.z}ejowski, M.\ 2003, A\&A, 406, 855 
\bibitem[\protect\citeauthoryear{Mor \& Netzer}{2012}]{Mor12} Mor, R., Netzer, H.~2012, MNRAS, 420, 526-541
\bibitem[\protect\citeauthoryear{Nalewajko \ea}{2011}]{Nal11} Nalewajko, K., Giannios, D., Begelman, M. C., \ea~2011, MNRAS, 413, 333-346
\bibitem[\protect\citeauthoryear{Nalewajko \& Begelman}{2012}]{Nal12} Nalewajko, K., \& Begelman, M. C., \ea~2012, 427, 2480-2486
\bibitem[\protect\citeauthoryear{Nesci}{2014}]{ATel6542} Nesci, R.~2014,  ATel \#6542
\bibitem[\protect\citeauthoryear{Ogle \ea}{2005}]{Ogle05} Ogle, P. M., Davis, S. W., Antonucci, R. R. J., \ea~2005, ApJ, 618, 139-154
\bibitem[\protect\citeauthoryear{Poutanen \& Stern}{2010}]{Pou10} Poutanen, J. \& Stern, B.,~2010, ApJ, 717, L118-L121
\bibitem[\protect\citeauthoryear{Pozo Nu{\~n}ez \ea}{2012}]{Pozo12} Pozo Nu{\~n}ez, F., Ramolla, M., Westhues, C. \ea 2012, A\&A, 545, A84
\bibitem[\protect\citeauthoryear{Pozo Nu{\~n}ez \ea}{2014}]{Pozo14} Pozo Nu{\~n}ez, F., Haas, M., Ramolla, M. \ea 2014, A\&A, 568, A36
\bibitem[\protect\citeauthoryear{Peterson \ea}{2004}]{Pet04} Peterson, B. M., Ferrarese, L., Gilbert, K. M. \ea 2004, ApJ, 613, 682-699
\bibitem[\protect\citeauthoryear{Pushkarev \ea}{2009}]{Push09} Pushkarev, A. B., Kovalev, Y. Y., Lister, M. L., \ea~2009, A\&A, 507, L33 
\bibitem[\protect\citeauthoryear{Richards \ea}{2006}]{Rich06} Richards, G. T., Lacy, M., Storrie-Lombardi, L. J., \ea~2006, ApJS, 166, 470-497
\bibitem[\protect\citeauthoryear{Shang \ea}{2011}]{Sha11} Shang, Z., Brotherton, M. S., Wills, B. J., \ea~2011, ApJS, 196, 2
\bibitem[\protect\citeauthoryear{Sikora \ea}{1994}]{Sik94} Sikora, M., Begelman, M. C., \& Rees, M. J., 1994, ApJ, 421, 153
\bibitem[\protect\citeauthoryear{Sikora \ea}{2009}]{Sik09} Sikora, M., Stawarz, {\L}., Moderski, R., \ea~2009, ApJ, 704, 38
\bibitem[\protect\citeauthoryear{Sikora \ea}{2013}]{Sik13} Sikora, M., Janiak, M., Nalewajko, K., \ea~2013, ApJ, 779, 68 
\bibitem[\protect\citeauthoryear{Tanaka \ea}{2014}]{Tan14atel} Tanaka, Y.~T., Doi, A., Inoue, Y., \ea~2014, ATel \#6529
\bibitem[\protect\citeauthoryear{Tanaka \ea}{2015}]{Tan15} Tanaka, Y.~T., Doi, A., Inoue, Y., \ea~2015, ApJL, 799, L18
\bibitem[\protect\citeauthoryear{Tavecchio \& Ghisellini}{2008}]{Tav08} Tavecchio, F., \& Ghisellini, G., 2008, MNRAS, 385, L98-L102
\bibitem[\protect\citeauthoryear{Walker \ea}{2015}]{Wal87} Walker,~R.~C., Benson,~J.~M., \& Unwin,~S.~C.,~1987, ApJ, 316, 546
\bibitem[\protect\citeauthoryear{Zdziarski \& Grandi}{2001}]{Zdzia01} Zdziarski,~A.~A., \& Grandi,~P.,~2001, ApJ, 551, 186-196
\end{thebibliography}
\end{document}